\documentclass{vgtc}                          

\ifpdf
  \pdfoutput=1\relax                   
  \usepackage{graphicx}                
  \DeclareGraphicsExtensions{.pdf,.png,.jpg,.jpeg} 
\else
  \usepackage{graphicx}                
  \DeclareGraphicsExtensions{.eps}     
\fi%

\graphicspath{{figures/}{pictures/}{images/}{./}} 

\usepackage[para]{threeparttable}      
\usepackage{microtype}                 
\PassOptionsToPackage{warn}{textcomp}  
\usepackage{textcomp}                  
\usepackage{mathptmx}                  
\usepackage{times}                     
\usepackage{cite}                      
\usepackage{tabu}                      
\usepackage{booktabs}                  
\usepackage{xurl}
\usepackage{breakurl}
\usepackage{enumitem}
\usepackage{fancyhdr}
\usepackage{lipsum}

\onlineid{0}

\vgtccategory{Research}

\fancypagestyle{specialfooter}{%
  \fancyhf{}
  
  \fancyfoot[L]{\footnotesize This paper has been peer-reviewed and accepted to \textbf{VisActivities: 2nd IEEE VIS Workshop on Data Vis Activities to Facilitate Learning, Reflecting, Discussing, and Designing}, held in conjunction with IEEE VIS 2021, New Orleans, LA, USA. 2021 - Workshop organizers: Samuel Huron, Benjamin Bach, Georgia Panagiotidou, Mandy Keck, Jonathan C. Roberts, Sheelagh Carpendale, http://visactivities.github.io.}
}
\title{Experience of Teaching Data Visualization using Project-based Learning}

\author{Dietrich Kammer\thanks{e-mail: kammer@htw-dresden.de}\\ %
        \scriptsize University of Applied Sciences Dresden %
\and Elena Stoll\thanks{e-mail: elena.stoll@htw-dresden.de}\\ %
     \scriptsize University of Applied Sciences Dresden
\and Adam Urban\thanks{e-mail: adam.urban@htw-dresden.de}\\ %
     \scriptsize University of Applied Sciences Dresden}

\teaser{
  \centering
   \includegraphics[width=0.8\linewidth]{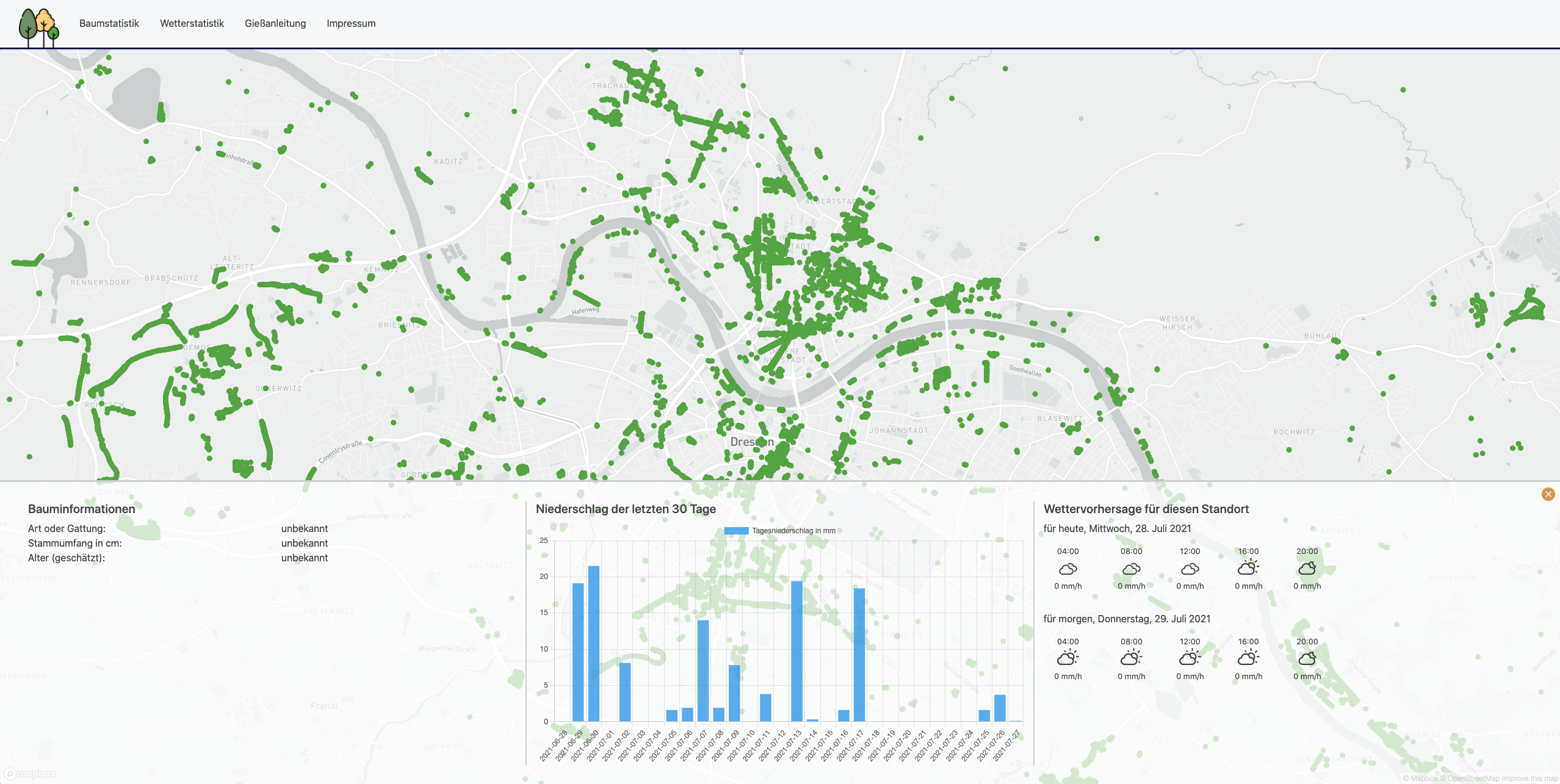}
  \caption{Student application based on JavaScript, D3, and Mapbox that shows trees in the neighbourhood that require watering -- a result from a course on visualizing urban sensor data.}
  \label{fig:teaser}
}

\abstract{We report our experience in two installations of a course on data visualization that featured project-based learning. Given the rationale of this approach, we show which input was provided when necessary for the students to achieve their goals. We also discuss and compare the tools we found useful for students to accomplish their goals. Fitting project-based learning into the standard schedule of a semester at University is a particular challenge, because it is hard for students to devote longer periods of time when working on their projects. Furthermore, online learning was a challenge to effectively perform group activities and work. We address didactic considerations, our course structure, tools, and student projects. Finally, we draw conclusions on the results and improvements in the course structure.%
} 

\CCScatlist{
  \CCScatTwelve{Human-centered computing}{Visualization}{}{};
  \CCScatTwelve{Social and professional topics}{Computing education}{Computing education programs}{Computer science education};
  \CCScatTwelve{Social and professional topics}{Computing education}{Student assessment}{}
}

\keywords{Visualization Education, Project-based Learning, Motivation, Visualization Literacy}

\begin{document}



\maketitle

\thispagestyle{specialfooter}%

\section{Introduction} 

Every scholar who teaches the basic theory, methods, and tools to create data visualizations needs to carefully select and adapt existing teaching material. Over the past years, a wealth of research \cite{Borner2019, godwin2016letsplay, beyer2016teaching}, several text books \cite{munzner2014visualization, riche2018datadrivenstorytelling, brehmer2021datasketches}, data visualization activities \cite{VisKit, wang2019teaching, huron2020ieee}, and online resources \cite{marketplace, visualizingdata} have emerged. While a teacher might have a broad overview about the field and in-depth insights into specific challenges and problems, laymen can be easily overwhelmed with the diverse aspects involved in creating a data visualization. 
In this paper, we show how we applied project-based learning to tackle this issue. A key idea of project-based learning is to encourage active participation of learners by means of developing a solution for a real-world problem over a longer period of time. 
It is particularly challenging to fit project-based learning into a standard semester schedule at University. The time allocated for one course is limited and work on a project is disrupted by other courses throughout the week. We describe our experience in two installations of a course on data visualization. By discussing the methods and tools we used, we can show advantages and disadvantages. We hope to encourage other scholars to adopt project-based learning also when facing organizational restrictions and share their experiences and resources.

\section{Didactic Considerations}

Project-based learning (PBL) has been named a transformational learning practice for the 21st century \cite{learningshift}. In line with modern approaches to learning, such as active learning \cite{prince2004activelearning}, the focus is on the learners who engage actively to construct knowledge. The learning process is driven by a real-world problem for which a solution is being developed over a longer period of time, usually in groups. Due to the complexity, ambiguity, and uniqueness of real challenges, there are no ready-made, i.e., ``right'' solutions, but rather a whole range of context-specific solutions. Thus, learners practice generic and complex cognitive skills, such as critical evaluation, creativity, the ability to abstract, communication, or reflection on the learning process itself. To unlock the potential in practice, however, projects must encourage student motivation and thoughtfulness. Teachers should regard themselves as counselors and must be supported in adapting and implementing this type of instruction \cite{Blumenfeld:1991:MPB}.

Within the context of visualization education, concepts of PBL and active learning have been adapted and applied successfully. 
For example, PBL impacts and promotes student engagement, participation, and motivation in data analysis courses as shown by \cite{nunez2017dataanalytics}. This is consistent with the findings of \cite{syda2020designstudylite} who choose a problem-oriented approach similar to PBL by incorporating service-learning. Students in these data visualization courses were more dedicated and committed to their design study projects as they could potentially make a real impact. \cite{Cheng2019collaborativePBL} report a positive effect of collaborative PBL on students' interest and curiosity in data science. To promote visualization capacity building, \cite{byrd2016curriculum} revise the curriculum for data visualization courses, incorporating both PBL and active learning as integral components.

The ``project seminar'' in the bachelor curriculum for our students of media computer science suits PBL perfectly with the following learning objectives. Students should work scientifically and individually acquire knowledge about a specific topic. These topics can cover all areas of computer science and should not have been taught explicitly before. Students should learn to analyze complex challenges and problems and work on solutions in teams or alone. They should learn to document solutions in a scientific sound manner. Our grading scheme included points for motivation, work style, presentation of results, project scope, quality, sustainability, and documentation.

\begin{figure}[!htb]
 \centering 
 \includegraphics[width=\columnwidth]{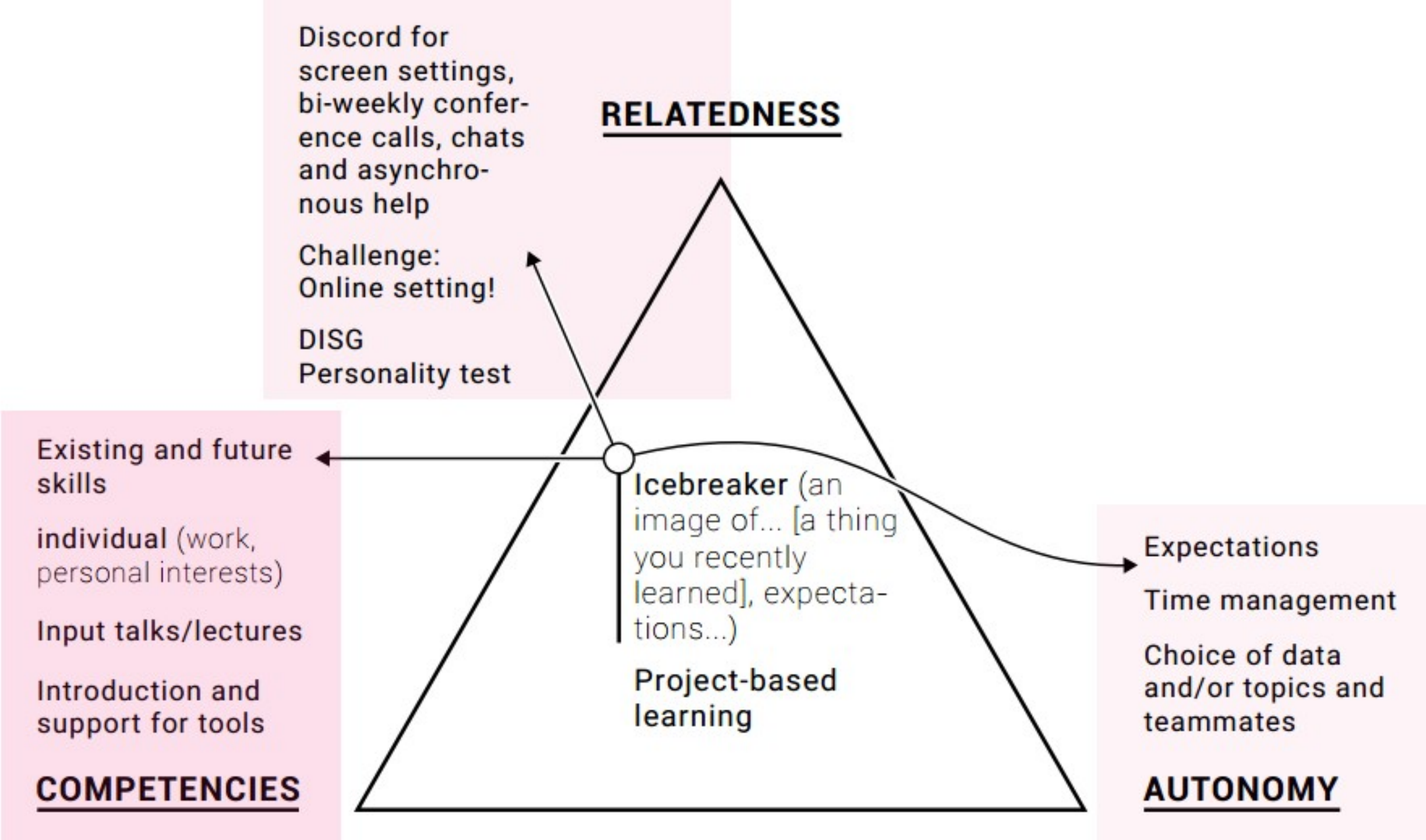}
 \caption{Exploration of motivational factors in our data visualization courses according to the Self-determination Theory.}
 \label{fig:sdt-v1}
\end{figure}

Within our PBL approach, we aimed to incorporate Deci and Ryan’s Self-determination Theory \cite{van_lange_self-determination_2012}. This theory suggests that autonomous motivation results from the fulfillment of three innate psychological needs: relatedness, autonomy, and competence. Recent findings indicate that relatedness is the strongest contributor to motivation \cite{wang2019competence}. We applied the theory to our courses on data visualization in order to investigate to what extent the three psychological needs were met (see Figure \ref{fig:sdt-v1}). Due to the COVID-19 pandemic the course was held online, which may have increased need for relatedness, but decreased the need for autonomy. Becoming aware of this shift in motivational needs throughout the first course, we decided to start the second course with an icebreaker activity. Participants and counselors introduced themselves with a picture, reported on something recently learned, formulated their expectations, and conducted a voluntary DISC \footnote{https://discpersonalitytesting.com/free-disc-test} personality test. By relating oneself to the others, a sense of the social structure and individual work styles emerged.

\section{Courses}

Both courses spanned the entire summer term (15 weeks) and involved three to four students of media computer science (see Table \ref{tab:courses}). The course was selected by the bachelor students as a mandatory elective module in their 4th or 6th semester. Hence, this voluntary decision shows a basic motivation or interest for the general theme of the course. Discord was used for online meetings and asynchronous communication as well as exchanging documents and references. 
Each course was planned with specific milestones to ensure continuous progress over the semester. The \textbf{A) preparatory phase} (4 weeks) included general research, specification of an application area, team-building, and data acquisition. The \textbf{B) data preparation phase} (3 weeks) was followed by the \textbf{C) concept phase} (3 weeks) and \textbf{D) implementation and documentation phase} (5 weeks) for developing a prototypical solution. Each phase presented specific challenges such as finding appropriate data sources (phase A), cleaning and preparing the data (B), conceiving and evaluating alternative interface and visualization concepts (C) as well as finding and using appropriate frameworks and libraries (D). To support solving these challenges in a self-determined way, counselors offered bi-weekly consultations hosting critiques on the work status and provided tutorials and guidance on demand.

In the first six weeks, we prepared accompanying input lectures for the students to facilitate team work and present basic methods in creating data visualizations:

\begin{enumerate}[noitemsep]
    \item \textbf{Team work}: Group processes (roles) and general collaboration tools (GitHub, Miro, etc.).
    \item \textbf{Introduction to data visualization}: Visualization Pipeline by Card et al., scope and applications of visualization, examples from history (static) and present (interactive).
    \item \textbf{Data Handling}: Syntax and semantics of data and classification (nominal, ordinal, quantitative), data acquisition, cleaning, and preparation.
    \item \textbf{Basic Visual principles}: Visual variables, perception, Gestalt laws, color use, scales.
    \item \textbf{Diagram types}: Examples and construction of diagrams for comparison, hierarchies, interdependence, graphs, tendencies, uncertainty, etc.
    \item \textbf{Tools}: Jupyter or ObservableHQ notebooks, respectively for each course, and Data Driven Documents (D3).
\end{enumerate}

Note that lectures on \textbf{User Centered Design} and \textbf{Interactivity in data visualization} were prepared but not used due to the limited time and other challenges faced by the students (see section \ref{sec:discussion}).

\begin{table}[tb]
  \caption[Overview]{Courses overview with autonomy referring to the freedom of choice regarding the concrete topic or application domains.}
  \label{tab:courses}
\begin{threeparttable}
  \scriptsize%
	\centering%
  \begin{tabular}{l l l}
  \toprule
    & \textbf{1st course} & \textbf{2nd course} \\
  \toprule
	\textbf{Students} & \multicolumn{2}{c}{\parbox{5.7cm}{\centering Mandatory elective module for 2nd or 4th semester bachelor students at University of Applied Sciences}}  \\ 
	\midrule
	\textbf{Prerequisites} & \multicolumn{2}{c}{\parbox{5.66cm}{\centering Previous knowledge in programming and computer science foundations needed}} \\
	\midrule
	\textbf{Conditions} & \multicolumn{2}{c}{\parbox{5.66cm}{\centering 15 weeks with a milestone plan and group or individual work}}  \\
	\midrule
    \textbf{Topic} & \parbox[t]{2.8cm}{\raggedright Visualization of Model Training in Natural Language Processing} &  \parbox[t]{2.8cm}{\raggedright Visualization of Urban Sensor Data} \\
    \midrule
    \textbf{Autonomy} & Medium & High \\
    \midrule
    \textbf{Groups} & \parbox{2.8cm}{\raggedright One group of 2 students, one individual project} & \parbox{2.8cm}{\raggedright One group of 2 students, two individual projects} \\
    \midrule
    \textbf{Technologies} & \parbox{2.8cm}{\raggedright Jupyter\footnotemark and Google Colaboratory\footnotemark Notebooks with Numpy\footnotemark, Pandas\footnotemark, Google BERT\footnotemark, Flair\footnotemark, Plotly\footnotemark, and Matplotlib\footnotemark} & \parbox{2.8cm}{\raggedright ObservableHQ\footnotemark Notebooks with D3\footnotemark, VegaLite\footnotemark, jQuery\footnotemark, Mapbox\footnotemark as well as Excel and Java for Data Processing} \\
    \midrule
    \textbf{Pass rate} & 100\% & 75\%  \\
  \bottomrule
  \tabuphantomline
  \end{tabular}
  \begin{tablenotes}[flushleft]
    \item[2] https://jupyter.org
    \item[3] https://colab.research.google.com
    \item[4] https://numpy.org
    \item[6] https://github.com/google-research/bert
    \item[5] https://pandas.pydata.org
    \item[7] https://github.com/flairNLP/flair
    \item[8] https://plotly.com
    \item[9] https://matplotlib.org
    \item[10] https://observablehq.com
    \item[11] https://d3js.org
    \item[12] https://vega.github.io/vega-lite
    \item[13] https://jquery.com
    \item[14] https://www.mapbox.com \hspace{10em}
  \end{tablenotes}
\end{threeparttable}
\end{table}

\subsection{Visualizing Model Training in Natural Language Processing}

The motivation for the first course was to explore means of data visualization in order to enhance the understanding of machine learning processes (Explainable AI), specifically Natural Language Processing. To this end, students first needed to acquaint themselves with Jupyter notebooks and two major libraries: Google BERT and the Flair NLP framework. We also offered to use Stanford NLP, which was quickly disregarded due to its inferior performance. Two students used the Flair framework in their project and another student BERT. 

\begin{figure}[!htb]
 \centering 
 \includegraphics[width=\columnwidth]{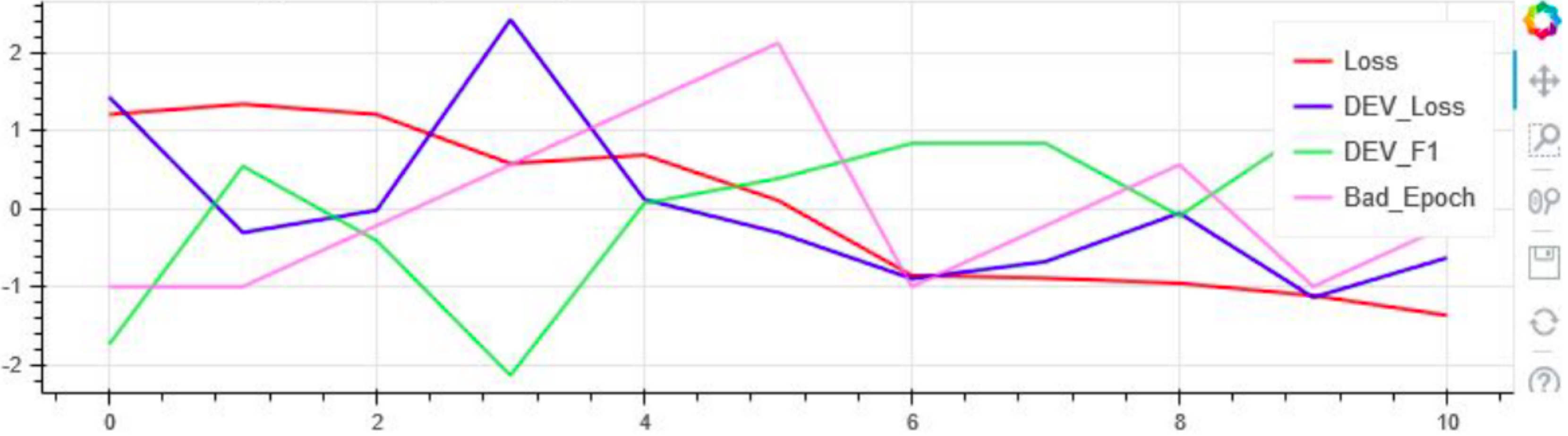}
 \caption{Visualization of the progress over multiple epochs (X-axis) in FLAIR model training (standardized with z-transformation for better comparison) using Seaborn library (based on Matplotlib).}
 \label{fig:progress}
\end{figure}

\begin{figure}[!htb]
 \centering 
 \includegraphics[width=\columnwidth]{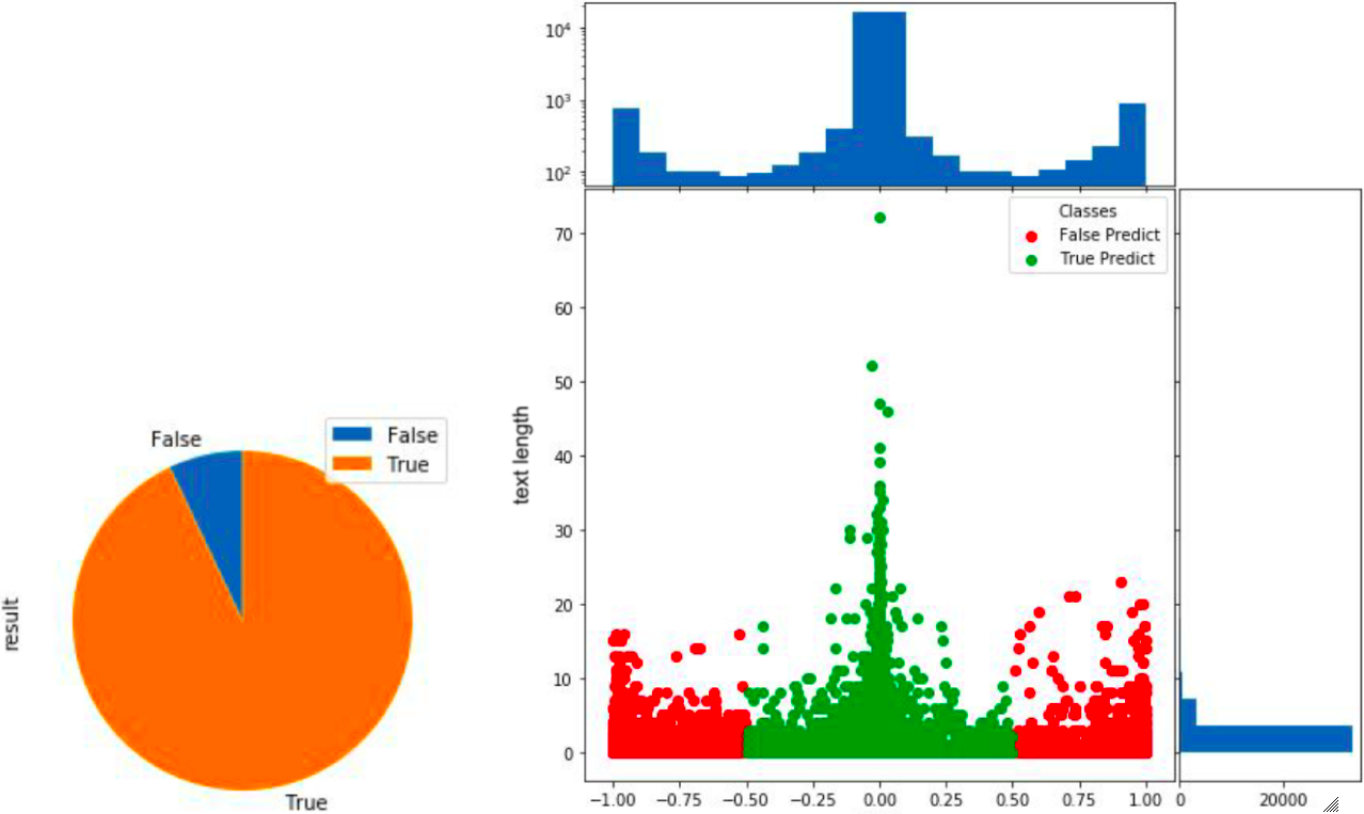}
 \caption{Visualizations that show false and true predictions after the NLP model has been trained using BERT (left) and how text length related to predictions (right) (using Matplotlib library).}
 \label{fig:textlengths}
\end{figure}

Students spent about 8 weeks setting up their working environment and acquiring basic Python skills along with learning the fundamental libraries for reading and manipulating data (Numpy, Pandas). Moreover, the basic concepts of NLP were researched such as word tokenization, word embeddings, named entity recognition, and so on. By running notebooks on Google's Colab platform, models could be trained efficiently. Although a lot of time was spent on such preliminary activities, all students managed to create at least one visualization (using Plotly, Matplotlib, and Seaborn libraries in the notebooks) that either show insights into the training progress (see Figure \ref{fig:progress}) or training data and classification quality (Figure \ref{fig:textlengths}). 

The level of autonomy in the first course was lower compared to the second. The choice of either of the three available NLP frameworks was a fixed requirement. The choice of data via publicly available datasets was also restricted since they had to be suitable for the model training and contain certain classifications. A set of tweets about an airline with an analysis of sentiments was selected and customer feedback from Yelp.

\subsection{Visualization of Urban Sensor Data}

The second course was intentionally kept more open with regard to the possible topics. The main motivation here was to review data that is collected in urban areas by different sensor networks or institutions. In contrast to the previous course, the freedom to choose a specific topic of interest was greater. We found that all students were highly motivated to work with environmentally significant data. A group of two students were interested in pollution and traffic statistics, while another student worked on an individual project with regard to watering trees in the city that are in danger of dehydration. A third student's desire was to retrieve an overview of historical buildings in the city. However, despite offering alternative directions and individual consultations, this student decided to quit after eight weeks and repeat the course later in his studies due to the difficulty of finding appropriate data from public sources. 

\begin{figure}[!htb]
 \centering 
 \includegraphics[width=\columnwidth]{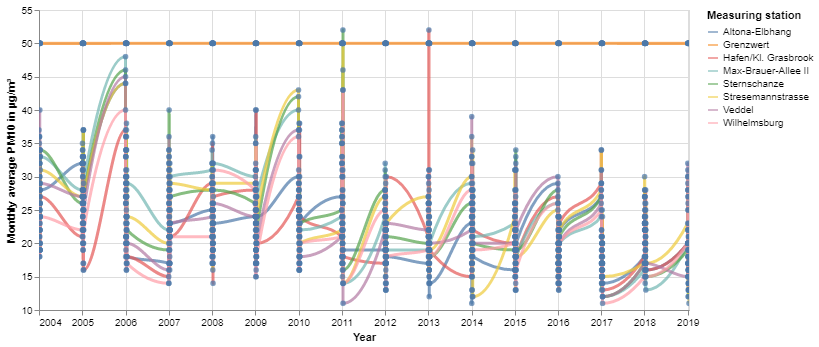}
 \caption{Showing the data from different measurement stations regarding the pollution of the air with PM10 fine dust particles using VegaLite in an ObservableHQ notebook.}
 \label{fig:PM10}
\end{figure}

ObservableHQ in conjunction with D3 and VegaLite was used to create initial insights into the data and create visualizations (see Figure \ref{fig:PM10}). The resulting stand-alone web applications were both based on Mapbox (see Figures \ref{fig:teaser}). 

\section{Discussion and Lessons Learnt}
\label{sec:discussion}

During our two project-based courses on data visualization, we identified several pain points. We monitored the students closely during the semester and took notes on their progress. Our final assessments included points for team work, frequency and quality of communication, reliability, and so on. In their project documentation, students were asked to write a section to personally reflect the course. Before announcing their grades, we explicitly invited each student to self-reflect their achievements and progress. Although almost all students were successful in creating and documenting a data visualization project, we could derive more possible actions from our observations to ensure that the teaching and learning process runs smoothly. 

\begin{itemize}[noitemsep]
    \item \textbf{Data is key}: In all projects, a major issue was the acquisition and preparation of data. This strong emphasis on the first two steps in the visualization pipeline led to a neglect of the issue of interactivity in data visualization. Data preparation is key challenge and very time consuming. 
    \item \textbf{Prior knowledge is key}: Prerequisites for the course were basic programming skills and knowledge in computer science foundations. Students with more coding skills were able to apply them to the problem of preparing data and implementing a visualization with greater ease. There is an understandable tendency to use already known technologies. This way, students can dedicate their resources to more complex tasks.
    \item \textbf{Hidden Problems}: We encountered several late questions regarding important issues in the project work. Most likely, the online setting worsened this issue. Students occasionally showed restraint to verbalize uncertainty and issues in contrast to a more personal setting where non-verbal cues are visible by the instructors. This shows shortcomings in the relatedness aspect of the motivation theory by Deci \& Ryan. 
    \item \textbf{Autonomy can lead to frustration}: One student quit the second course due to initial problems in finding data he was personally interested in. Although we suggested to look for alternative data or create artificial datasets, the determination to find specific data of personal interest stood in the way of completing a successful project.
    \item \textbf{Autonomy can lead to extraordinary results}: In contrast to the issue described above, autonomous planning can also lead to very good results. The most motivated students worked alone, spent far more time on the course than the curriculum called for, and produced extraordinarily creative, thoughtful, and comprehensive results. This is a testimony to the power of intrinsic motivation according to Deci \& Ryan. 
    \item \textbf{Scientific writing and documentation}: A general issue for all projects was the documentation. The quality was especially limited due to missing skills in scientific writing.
\end{itemize}

In our two courses, the degree of autonomy was slightly varied for choosing a topic. With a greater pass rate in the more restricted setting, it seems that this approach is more sound. However, the scope and quality of the projects in the second course also emphasizes that when properly accompanied, giving more freedom to choose a topic boosts personal involvement. Naturally, given the 7 students in total, we have to accept that individual predispositions and diversity also play an important role. In the following section, we present suggestions to improve project-based learning for teaching data visualization.

\section{Future Work}

With regards to the above issues, we plan mandatory individual consultations to obtain direct feedback during the semester. Several assessment techniques can be used and adapted, e.g. the submission of a ``muddiest point'' after each input lecture or consultation. Similarly, a minute paper written by the students could show their understanding of a topic or problem. With regards to team work, three questions can be asked repeatedly during the semester: How satisfied are you with your work and participation? How satisfied are you with the work and participation of your fellow students? How satisfied are you with the work of your instructors? This systematic satisfaction assessment can help to (anonymously) discover personal issues and monitor the teaching and learning process more closely. 

In addition to our existing grading scheme, we would like to formally assess the visualization literacy of the participating students, e.g. using tests such as \cite{Lee:2017:VLAT}. In our experience, interactivity in the developed applications was the most neglected feature. We assume that bigger teams with a specific focus of each team member (data preparation vs. interaction concept) could improve this situation. However, finding agreement on a mutually interesting topic is challenging. Alternatively, the interaction concept could be refined in a second course or project (e.g. bachelor thesis). Other possibilities are to balance the amount of inspiration for possible projects, provide more examples, or even rough scaffolds for programs that can be adapted to specific problems. Deadlines for the topic selection could be made more strict, forcing to stay with one topic even when data is not easiliy accessible. 
The project-based learning presented here can also greatly benefit from cooperation with organizations or companies outside of University, showing the real-world need for solutions to specific data problems. 

\acknowledgments{
The authors wish to thank the students participating in the courses.}

\bibliographystyle{abbrv-doi}

\bibliography{template}
\end{document}